\documentstyle[aps,prd,epsfig]{revtex}
\begin{document}
%%%%%%%%%%%%%%%%%%%%%%%%%%%%%%% title page %%%%%%%%%%%%%%%%%%%%%%%%%%%%%%
\draft
\title{Transverse Double-Spin Asymmetries 
for Muon Pair Production in $pp$ Collisions}
\author{O.~Martin, A.~Sch\"afer}
\address{Institut f\"ur Theoretische Physik, Universit\"at Regensburg,
D-93040~Regensburg, Germany}
\author{M. Stratmann}
\address{Department of Physics, University of Durham, Durham, DH1~3LE, England}
\author{W. Vogelsang}
\address{Theory Division, CERN, CH-1211~Geneva~23, Switzerland; now at: \\
Institute for Theoretical Physics, State Univ. of New York
at Stony Brook, NY-11794, USA}
\date{\today}
\maketitle
%

%%%%%%%%%%%%%%%%%%%%%%%%%%%%%%% abstract %%%%%%%%%%%%%%%%%%%%%%%%%%%%%%%%%%%
\begin{abstract}
We calculate the rapidity dependence of the transverse double-spin asymmetry 
for the Drell-Yan process to next-to-leading order in the strong coupling.
Input transversity distributions are obtained by saturating the Soffer
inequality at a low hadronic mass scale. Results for the polarized 
{\sc Bnl-Rhic} proton-proton collider and the proposed 
{\sc Hera}-$\mathrm\vec{\mbox{N}}$ fixed-target 
experiment are presented, and the influence of the limited muon
acceptance of the detectors on measurements of
the asymmetry is studied in detail. 
\end{abstract}
\pacs{}
\noindent
One of the major goals of the forthcoming spin programme at the {\sc Rhic}
polarized proton-proton collider \cite{rhic} is a first measurement of the
twist-2 transversity distribution $\delta q(x,\mu^2)$ \cite{trans},
which is theoretically as important as the well-known unpolarized and 
longitudinally polarized parton densities $q(x,\mu^2)$ and 
$\Delta q(x,\mu^2)$, respectively.
An important, non-trivial model-independent restriction on the size of
$\delta q(x,\mu^2)$ derives from Soffer's inequality \cite{soff}, which 
states that
\begin{equation}
|\delta q(x,\mu^2)| \leq \frac{1}{2} \left[ q(x,\mu^2) + \Delta q(x,\mu^2)
\right] \; ,
\label{soffersin}
\end{equation}
and similarly for antiquarks. 
It was shown to be preserved by next-to-leading order (NLO)
DGLAP evolution in ``reasonable'' factorization schemes, among them
the $\overline{\mathrm MS}$-scheme [4-6].

In a previous NLO analysis, see \cite{we} for details, we have derived 
an upper bound [by saturating (\ref{soffersin})] for the total transverse 
double-spin asymmetry $A_{TT}(M)$ for Drell-Yan dimuon production of mass
$M$. It turned out, however, that $A_{TT}(M)$ is not very sensitive to 
the {\em shape} of $\delta q$. 
In addition the angular acceptance of the detectors 
was assumed to be constant, i.e., independent of the dimuon rapidity $y$, 
which can only be a rather crude approximation to the real experimental 
conditions. Therefore, in order to better suit the experimental needs, we 
extend the analysis of \cite{we} in this note and study the $y$ dependence 
of $A_{TT}$.

The transversely polarized Drell-Yan cross section, 
${\mathrm d}\delta\sigma\equiv\left({\mathrm d}\sigma^{\uparrow\uparrow}-
{\mathrm d}\sigma^{\uparrow\downarrow}\right)/2$, is given as a double 
convolution of transversity distributions with the corresponding 
transversely polarized partonic cross section:
\begin{equation} 
\label{sig} 
\frac{{\mathrm d}\delta\sigma}{{\mathrm d}M{\mathrm d}y{\mathrm d}\phi}=
\sum_q \tilde{e}_q^2
\int_{x_1^0}^1 {\mathrm d}x_1 \int_{x_2^0}^1 {\mathrm d}x_2 \left[
\delta q(x_1,\mu_F^2) \delta\bar q(x_2,\mu_F^2) +
\delta\bar q(x_1,\mu_F^2) \delta q(x_2,\mu_F^2) \right]
\frac{{\mathrm
d}\delta\hat\sigma}{{\mathrm d}M{\mathrm d}y{\mathrm d}\phi} \;,
\end{equation}
$\mu_F$ being the factorization scale. 
The effective charge $\tilde{e}_q$ also contains the electroweak effects
from $Z^0$ exchange and $\gamma Z^0$ interference; 
see, e.g., Eq.~(20) of \cite{we}. 
$y$ denotes the rapidity of the dimuon pair, and $\phi$ is the azimuthal 
angle of one muon, with $\phi=0$ in the direction of positive transverse 
spin of the incoming protons. The variables $x_1^0$, $x_2^0$
in (\ref{sig}) are related to $y$ and the Drell-Yan scaling 
variable $\tau=M^2/S$ by $x_1^0 = \sqrt{\tau}e^y$ and 
$x_2^0=\sqrt{\tau}e^{-y}$, so that the region $y>0$ ($y<0$) is mainly 
sensitive to $x_1^0$ ($x_2^0$). 
To lowest order (LO) $x_1^0$ and $x_2^0$ coincide
with the momentum fractions carried by the incident partons. 
Indeed, one has at LO\footnote{The formula $\Phi(\phi)=1$ below
Eq.~(15) in \cite{we} should read $\Phi(\phi)=2$.}:
\begin{equation}
\label{eq:lo}
\frac{{\mathrm d}\delta\hat\sigma^{(0)}}{{\mathrm d}
M{\mathrm d}y{\mathrm d}\phi} = 
\frac{2\alpha^2}{9SM}\cos (2\phi) \delta(x_1-x_1^0)
\delta(x_2-x_2^0) \, .
\end{equation}
In the $\overline{\mathrm MS}$-scheme, the NLO [${\cal O}(\alpha_s)$] 
correction to (\ref{eq:lo}) reads
\begin{eqnarray}
\label{sigmanlo}
\frac{{\mathrm d} \delta \hat \sigma^{(1),
{\overline{{\mathrm MS}}}}}{{\mathrm d}
M{\mathrm d}y{\mathrm d}\phi} &=&
\frac{2\alpha^2}{9SM} C_F \frac{\alpha_s(\mu_R^2)}{2\pi}
\frac{4\tau(x_1x_2+\tau)}{x_1x_2(x_1+x_1^0)(x_2+x_2^0)}\cos (2\phi)
\nonumber \\
&\times& \left\{ 
\delta(x_1-x_1^0)\delta(x_2-x_2^0) \left[
\frac{1}{4}\ln^2\frac{(1-x_1^0)(1-x_2^0)}{\tau}
+\frac{\pi^2}{4}-2 \right]
\right. \nonumber \\ && \left. 
+ \delta(x_1-x_1^0) \left[ \frac{1}{(x_2-x_2^0)_{+}}
\ln\frac{2x_2(1-x_1^0)}{\tau(x_2+x_2^0)}
+\left(\frac{\ln(x_2-x_2^0)}{x_2-x_2^0}\right)_{+}
+\frac{1}{x_2-x_2^0}\ln\frac{x_2^0}{x_2}\right]
\right. \nonumber \\ && \left. 
+ \frac{1}{2[(x_1-x_1^0)(x_2-x_2^0)]_{+}}
+ \frac{(x_1+x_1^0)(x_2+x_2^0)}{(x_1 x_2^0+x_2x_1^0)^2}
- \frac{3\ln\left(\frac{x_1x_2+\tau}{x_1x_2^0+x_2x_1^0} \right)}{(x_1-x_1^0)
(x_2-x_2^0)}
\right. \nonumber  \\ && \left. 
+ \ln \frac{M^2}{\mu_F^2} \left[
\delta(x_1-x_1^0)\delta(x_2-x_2^0)\left(
\frac{3}{4}+\frac{1}{2}\ln \frac{(1-x_1^0)(1-x_2^0)}{\tau} \right) 
+\delta(x_1-x_1^0)\frac{1}{(x_2-x_2^0)_{+}} \right] \right\}
+ [1 \leftrightarrow 2] \, ,
\end{eqnarray}
where $\mu_R$ is the renormalization scale 
(for simplicity we always take $\mu_R \equiv \mu_F=M$) 
and $(i=1,\,2)$
\begin{equation}
\int_{x_i^0}^1 d x_i f(x_i) \frac{1}{(x_i-x_i^0 )_+} \equiv
\int_{x_i^0}^1 d x_i \frac{f(x_i)-f(x_i^0)}{x_i-x_i^0} \, .
\end{equation}
Equation~(\ref{sigmanlo}) is obtained by a suitable factorization-scheme 
transformation of the corresponding result of \cite{vowe}, which 
was calculated taking the gluon off-shell in the process $q\bar{q}\rightarrow 
\mu^+ \mu^- g$ in order to regularize its collinear divergences.
The corresponding results for the unpolarized NLO cross section 
${\mathrm d}\sigma\equiv\left({\mathrm d}\sigma^{\uparrow\uparrow}+ 
{\mathrm d}\sigma^{\uparrow\downarrow}\right)/2$ can be found in 
\cite{sutton}\footnote{The result of \cite{sutton} is not given in the conventional 
$\overline{\mathrm MS}$-scheme; however the translation can be easily made.}.
 
In order to increase the observable rates, we will integrate the unpolarized
cross section over $\phi$, whereas in the polarized case we add each
quadrant with a different sign. Thus, the rapidity dependent 
asymmetry will be defined as
\begin{equation}
A_{TT}(y) \equiv 
\frac{\int_{M_0}^{M_1}{\mathrm d}M 
\left(\int_{-\pi/4}^{\pi/4}-\int_{\pi/4}^{3\pi/4}+ 
\int_{3\pi/4}^{5\pi/4}-\int_{5\pi/4}^{7\pi/4} \right){\mathrm d}\phi
\,\,{\mathrm d}\delta\sigma/{\mathrm d}M{\mathrm d}y{\mathrm d}\phi}
{\int_{M_0}^{M_1}{\mathrm d}M\int_0^{2\pi}{\mathrm d}\phi 
\,\,{\mathrm d}\sigma/{\mathrm
d}M {\mathrm d}y {\mathrm d}\phi}\, ,
\label{att}
\end{equation}
where $M_{0,1}$ denote the limits of some suitable bin in invariant mass.
Following closely our previous study \cite{we} on the total 
(i.e.\ $y$-integrated) Drell-Yan cross section, we will try to 
estimate {\em upper bounds} on $A_{TT}$ by assuming that the {\em equality} 
in (\ref{soffersin}) holds\footnote{In \cite{we}
we actually did not saturate the {\em total} quark distributions, but only 
their {\em valence} component at the input scale $\mu_0$. As was pointed 
out in \cite{kana}, this is, strictly speaking, not the statement of the 
Soffer inequality. A careful numerical check however reveals that none of our 
results in \cite{we} is altered if one saturates the {\em full} quark distributions 
instead of the valence ones.} at a low hadronic 
mass scale $\mu_0\simeq{\cal{O}}(0.6\,\mathrm{GeV})$, see \cite{we} for more
details.
We should emphasize that the {\em sign} of the 
asymmetry cannot be predicted in this way, because only the absolute 
value of $\delta q$ enters Soffer's inequality. This also means 
that all possible combinations
of signs in Eq.~(\ref{soffersin}) must be checked so as to obtain the 
maximal absolute value of $A_{TT}$. In our case, choosing all signs to 
be the same always yielded the largest results.

Figures~\ref{fig1} and \ref{fig2} show the ``maximally possible'' 
$d\delta\sigma/dy$ and $A_{TT}$ in LO and NLO for
$\sqrt{S}=500$~GeV at {\sc Rhic} and for $E_{\mathrm beam}=820$~GeV,
corresponding to $\sqrt{S}=39.2\,{\mathrm{GeV}}$, at 
{\sc Hera}-$\vec{\mathrm N}$, respectively. 
We have integrated over $M$ in (\ref{att})
as indicated in the figures, avoiding masses smaller than
5~(4)~GeV for {\sc Rhic} ({\sc Hera}-$\vec{\mathrm{N}}$),
where a large background from charmed-meson decay is expected. 
Very similar results as in Fig.~\ref{fig1} are obtained for
$\sqrt{S}=200\,\mathrm{GeV}$ and ${\cal{L}}=320\,\mathrm{pb}^{-1}$ at
{\sc Rhic} when restricting $M$ to be in the range $5-9\,\mathrm{GeV}$.
The QCD corrections to the 
polarized cross section turn out to be largest in the fixed-target regime, 
whereas the asymmetry receives the largest corrections at
higher energies. In most cases the NLO contributions
are sizeable and should be included for a meaningful comparison with 
future data. We note in passing that we found that the dependence of 
the results on $\mu_R$ and $\mu_F$ is greatly reduced at NLO. 

In Figs.~\ref{fig1} and \ref{fig2} we also display the statistical errors 
expected for such measurements of $A_{TT}$. Here, we try
to estimate the influence of detector cuts on the error, which could be
rather crucial for making realistic predictions. For instance, if the
muon detectors have limited angular coverage, one or both of the muons
might escape detection just for geometrical reasons, and the event is lost. 
In the case of the {\sc Rhic} detector {\sc Phenix}\footnote{We only calculate 
acceptance corrections for {\sc Phenix}, since the other major 
{\sc Rhic} detector, {\sc Star}, cannot detect muons, but only 
electrons. Electron pair production does not seem as promising as muon pair 
production, as a very detailed study of the background is required in that
case.}, the endcaps will be able to identify muons with $1.2<|y_{\mu^{\pm}}|
<2.4$; an additional cut on the muon momentum, $|\vec k|>2$~GeV, will 
probably be necessary to get rid of unwanted background. Central rapidity 
muon detector arms, which would cover $|y_{\mu^{\pm}}| < 0.35$ (even though
for only half of the azimuth), were proposed but will not be realized 
\cite{saito}. Nevertheless, we have also studied the impact that 
they would have had
on the achievable experimental accuracy. In order to calculate 
the relevant acceptances, the momenta of the outgoing muons must be known.
However, they cannot be reconstructed from the kinematic variables
$M$, $y$, and $\phi$ introduced above, since $M$ and $y$ refer to the
dimuon system and $\phi$ is only one of the angles describing the direction 
of one muon. Therefore, one has to consider a more differential cross section, 
like
\begin{equation}
\label{ktxsec}
\frac{{\mathrm d}(\delta)\hat\sigma^{(0)}}{{\mathrm d}
M{\mathrm d}y{\mathrm d}\phi{\mathrm d}k_T}
= \frac{4\alpha^2}{3SM^3}\frac{(\delta)\zeta(M,k_T,\phi)}
{\sqrt{1-\frac{4k_T^2}{M^2}}}
\delta(x_1-x_1^0) \delta(x_2-x_2^0) \, ,
\end{equation}
where $\zeta(M,k_T,\phi)=k_T\left(2-4 k_T^2/M^2\right)$, 
$\delta \zeta(M,k_T,\phi)=4\cos (2 \phi)k_T^3/M^2$, 
and $k_T$ is the transverse momentum of the muons. The LO acceptance 
curve for the measurement of, say, the $y$-dependence of the cross section or
the asymmetry $A_{TT}$, can then be obtained by dividing the results based 
on Eq.~(\ref{ktxsec}), after implementation of 
appropriate cuts on $y_{\mu^{\pm}}$,
by the full LO result, i.e., the one integrated over all $k_T$ and already 
used in Figs.~\ref{fig1} and \ref{fig2}. 
Of course one could have also extended the acceptance analysis to NLO, 
where the muons are no longer back-to-back and the possibility arises 
that both muons go into the same hemisphere of the detector. 
However, we believe that a LO estimate for the acceptance is good enough 
to get a rough quantitative understanding of the influence of limited detector
coverage on the statistical error. 

Figure~\ref{fig3} shows the acceptances for muon identification in the
endcaps only and for the endcaps plus central detector arms.
Note that the unpolarized acceptances $\varepsilon$ differ from 
the polarized ones $\delta\varepsilon$ as a result of the 
different $k_T$-dependences of the corresponding 
cross sections (\ref{ktxsec}). The results for 
$\sqrt{S}=200$~GeV and 500~GeV turn out to be almost the same, 
because we used the same lower limit for the dimuon mass $M$ in 
both cases. According to Fig.~\ref{fig3}, the acceptance for the 
central rapidity region $y\approx 0$, where each endcap or each central 
arm detects one muon, is considerably smaller than for the large 
rapidity region, where both muons hit the same endcap. Also,
the ratio of ``polarized-to-unpolarized acceptance'' is smaller
than unity in the former case and larger than unity for the latter. 
This means that the experimentally measured asymmetry will be smaller 
at $y\approx 0$, but somewhat enhanced at large $y$ as compared to the 
values given in Fig.~\ref{fig1}. We also see that 
the addition of muon identification in the central arms would yield a much 
larger acceptance at small and intermediate dimuon rapidities than found
for the ``endcaps only'' scenario.

At the moment, {\sc Hera}-$\vec{\mbox{N}}$ only has the status of 
a fairly general proposal for a fixed-target $pp$ spin 
experiment at {\sc Hera} \cite{nowak}. Thus, 
nothing specific is known yet about appropriate kinematical cuts. In our
analysis we try to use reasonable values for the kinematical coverage, 
keeping in mind that the true detector could look significantly different 
in case it will ever be built. We use $\pm$700~mrad for the horizontal 
and $\pm$160~mrad for the vertical opening angle, while the beam pipe 
is assumed to cover $\pm$10~mrad. Such a detector would have much larger 
acceptances than {\sc Phenix}, as can also be seen in Fig.~\ref{fig3}.

Exploiting our LO estimates of the acceptances
$\varepsilon$ and $\delta \varepsilon$, we are now in a position
to calculate the expected statistical errors on the asymmetry. Here
we assume that it makes sense to adopt our LO acceptances curve also 
for the NLO calculation; see our discussion above. The
statistical error of the ``measured'' asymmetry, i.e., {\em after} correction 
for acceptance, is then just given by 
$1/{\cal P}^{2}\sqrt{{\cal L}\int\varepsilon{\mathrm d}\sigma}$
where ${\cal P}$ denotes the degree of polarization of each beam, ${\cal L}$
is the integrated luminosity, and
the integration goes over the bin under consideration. In order to consistently
match the error bars to Figs.~\ref{fig1} and \ref{fig2}, we obviously have to
weigh them by the ratio $\frac{\int {\mathrm d}\delta \sigma}{\int {\mathrm d}
\sigma} \; / \; \frac{\int \delta\varepsilon{\mathrm d} \delta\sigma}
{\int\varepsilon{\mathrm d}\sigma}$.

The statistical errors show the same features for both {\sc Rhic} energies. 
A measurement in the central rapidity region will hardly be possible, even
if the central muon detector arms are added. Statistical errors at large 
rapidities do not depend on the presence of central rapidity muon detection 
(see Fig.~\ref{fig3}), and prospects look slightly better here. The larger 
rates for $\sqrt{S}=500$~GeV are compensated by a smaller asymmetry so 
that, for both $\sqrt{S}=200$~GeV and $\sqrt{S}=500$~GeV, 
the relative statistical error is about 40\% at
large $y$. Note that we also include the events with negative rapidity 
for the calculation of the error bars, since the results are symmetric in 
$y$. The situation for {\sc Hera}-$\vec{\mbox{N}}$ is somewhat
better, with relative errors of about 30\%, and more possible bins. 
This is mainly due to the much larger asymmetry in the fixed-target regime. 
However, for all this we should keep in mind that the asymmetries we 
show have been obtained assuming saturated $\delta q$'s at a low scale. 
If the saturation were only at, say, the 50\% level, then all asymmetries 
would have to be scaled down by a factor $4$, and no measurement would
be possible.

Clearly, the restriction in angular acceptance expressed by Fig.~\ref{fig3}
will also leave its footprint for the $y$-integrated, i.e., the total, 
Drell-Yan cross section. In other words, we have to reinspect our
predictions made in \cite{we} for this quantity, to see whether 
there is any dramatic change concerning the statistical accuracy
of a possible measurement of $A_{TT} (M)$. On the left-hand sides of
Figs.~\ref{fig4} and \ref{fig6} we show the unpolarized and polarized 
acceptances for the total dimuon cross section for the {\sc Rhic}
energy of $\sqrt{S}=200$~GeV and the {\sc Hera}-$\vec{\mbox{N}}$ situation. 
In the case of {\sc Rhic},
we distinguish again between the ``endcaps only'' and the ``endcaps plus
arms'' options. The general trend is that the acceptances are rather low
for {\sc Rhic} ({\sc Phenix}) and decrease with increasing $M$ after
reaching a peak at a quite low $M$-value. Under our
assumed conditions for {\sc Hera}-$\vec{\mbox{N}}$, the acceptance turns
out to be much higher and fairly independent of $M$.
On the right-hand sides of 
Figs.~\ref{fig4} and \ref{fig6} we redisplay our findings for $A_{TT}(M)$ 
of Figs.~3 and 4 of Ref.~\cite{we}, but now with the more realistic error
bars based on our considerations concerning the acceptance. One finds that
at not too large $M$, a measurement of a non-vanishing asymmetry for the 
total Drell-Yan cross section still looks possible also for {\sc Rhic},
provided the ``true'' transversity densities are anywhere near 
the ones we have modeled. Measurements at large $M$ appear hopeless.
The situation for $\sqrt{S}=500$~GeV at {\sc Rhic} is qualitatively very similar
and hence not shown.
Again, as in the case of $A_{TT}(y)$, {\sc Hera}-$\vec{\mbox{N}}$ looks 
in a somewhat better shape.

In conclusion, we have studied the ``maximally possible'' 
$A_{TT}$, resulting from saturation of Soffer's 
inequality at a low hadronic scale. It turns out that the limited muon 
acceptance for the {\sc Rhic} experiments threatens to make a 
measurement of transversity elusive. In particular, it will be difficult, 
if not impossible, to measure the rapidity dependence of $A_{TT}$, which in 
principle would be expected to be sensitive to the {\em shape} of 
$\delta q$. At best, one data point at large $y$ can be 
obtained, but with a large relative error. 
The limitation in the muon acceptance also affects the $y$-integrated
cross section, so that the resulting $A_{TT}(M)$ will also receive a
substantial relative statistical error.
An upgrade of the {\sc Phenix} detector towards muon 
identification also in the central arms would not improve the situation 
significantly. Lower energies, in combination with better muon acceptance,
seem more favorable.
%
%%%%%%%%%%%%%%%%%%%%%%%%%%%%%%%%%%%%%%%%%%%%%%%%%%%%%%%
%
\section*{Acknowledgments}
We thank G.~Bunce and N.~Saito for useful information concerning 
the {\sc Rhic} detector acceptances. Furthermore, we are indebted
to W.-D.~Nowak for very valuable discussions about 
{\sc Hera}-$\vec{\mathrm{N}}$.
W.V. is grateful to D.\ de Florian 
for useful discussions. O.M. and A.S. acknowledge financial 
support from the BMBF and the ``Deutsche Forschungsgemeinschaft''. 
This work was supported in part by the EU Fourth Framework Programme 
``Training and Mobility of Researchers'', Network
``Quantum Chromodynamics and the Deep Structure of Elementary Particles'', 
contract FMRX-CT98-0194 (DG 12 - MIHT). 
%%%%%%%%%%%%%%%%%%%%%%%%%%%%%%%%%%%%%%%%%%%%%%%%%%%%%%%

%
%%%%%%%%%%%%%%%%%%%%%%%%%%
\section*{Figure Captions}
%%%%%%%%%%%%%%%%%%%%%%%%%%
%
\newcounter{fig}
\begin{list}{\bf Fig.~\arabic{fig}}{\usecounter{fig}}
\item \label{fig1} ``Maximal'' polarized cross section and 
asymmetry as functions of dimuon
rapidity $y$ for {\sc Rhic} at $\sqrt{S}=500\,{\mathrm GeV}$. The error bars 
have been calculated for ${\cal L}=800\,{\mathrm pb}^{-1}$, 70\% 
polarization of both beams, and include acceptance corrections (see text). 
The point at low rapidity can only be obtained if {\sc Phenix} is endowed 
with central muon detector arms.
\item \label{fig2} Same as Fig.~1, but for {\sc Hera}-$\vec{\mbox{N}}$ with 
$E_{\mathrm{beam}}=820\,{\mathrm GeV}$ and ${\cal L}=240\,\mathrm pb^{-1}$.
\item \label{fig3} Acceptance curves for the detection of dimuons with 
the {\sc Phenix} and {\sc Hera}-$\vec{\mbox{N}}$ detectors, as functions of
the dimuon rapidity $y$. The {\sc Phenix} acceptances for $\sqrt{S}=500$~GeV 
and $M=5$--$20$~GeV differ only very slightly from the results shown here
for the case $\sqrt{S}=200$~GeV.
\item \label{fig4} Dependence of the acceptances and the NLO 
asymmetry $A_{TT}$ on the dimuon
invariant mass, integrated over rapidity, for $\sqrt{S}=200$~GeV at 
{\sc Rhic}. The error bars on the right-hand side include the acceptance 
corrections and are based on ${\cal L}=320\,{\mathrm pb}^{-1}$ and ${\cal P}=0.7$. 
The outer error bars correspond to the ``endcaps only'' option, while 
the inner ones have been obtained assuming additional central detector arms.
%
%\item \label{fig5} Same as Fig.~4, 
%but for the parameters used in Fig.~\ref{fig1}. The error bars for 
%the bin $M>40$~GeV are much larger than the asymmetry and are therefore not 
%shown in the figure.
%
\item \label{fig6}
Same as Fig.~\ref{fig4}, but for $\sqrt{S}=39.2\,{\mathrm{GeV}}$, 
corresponding to {\sc Hera}-$\vec{\mbox{N}}$.
\end{list}
%
%%%%%%%%%%%%%%%%
%\newpage
%
\begin{center}

%\epsfig{file=fig1.eps,width=15cm}
%\large{\bf{Fig.~\ref{fig1}}}

\vspace*{3.8cm}
\epsfig{file=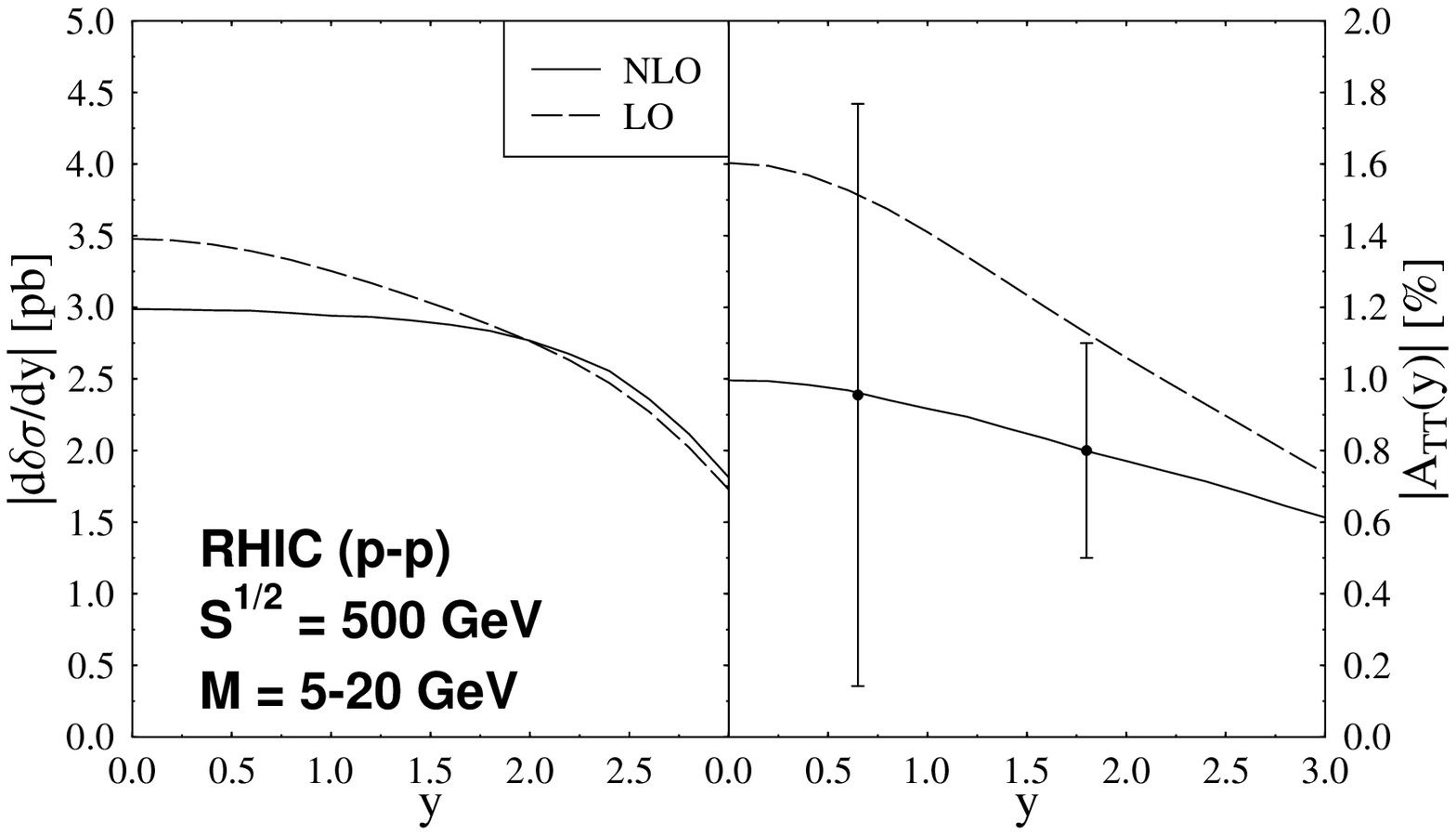,width=15cm}
\large{\bf{Fig.~\ref{fig1}}}

%\vspace*{0.8cm}
\epsfig{file=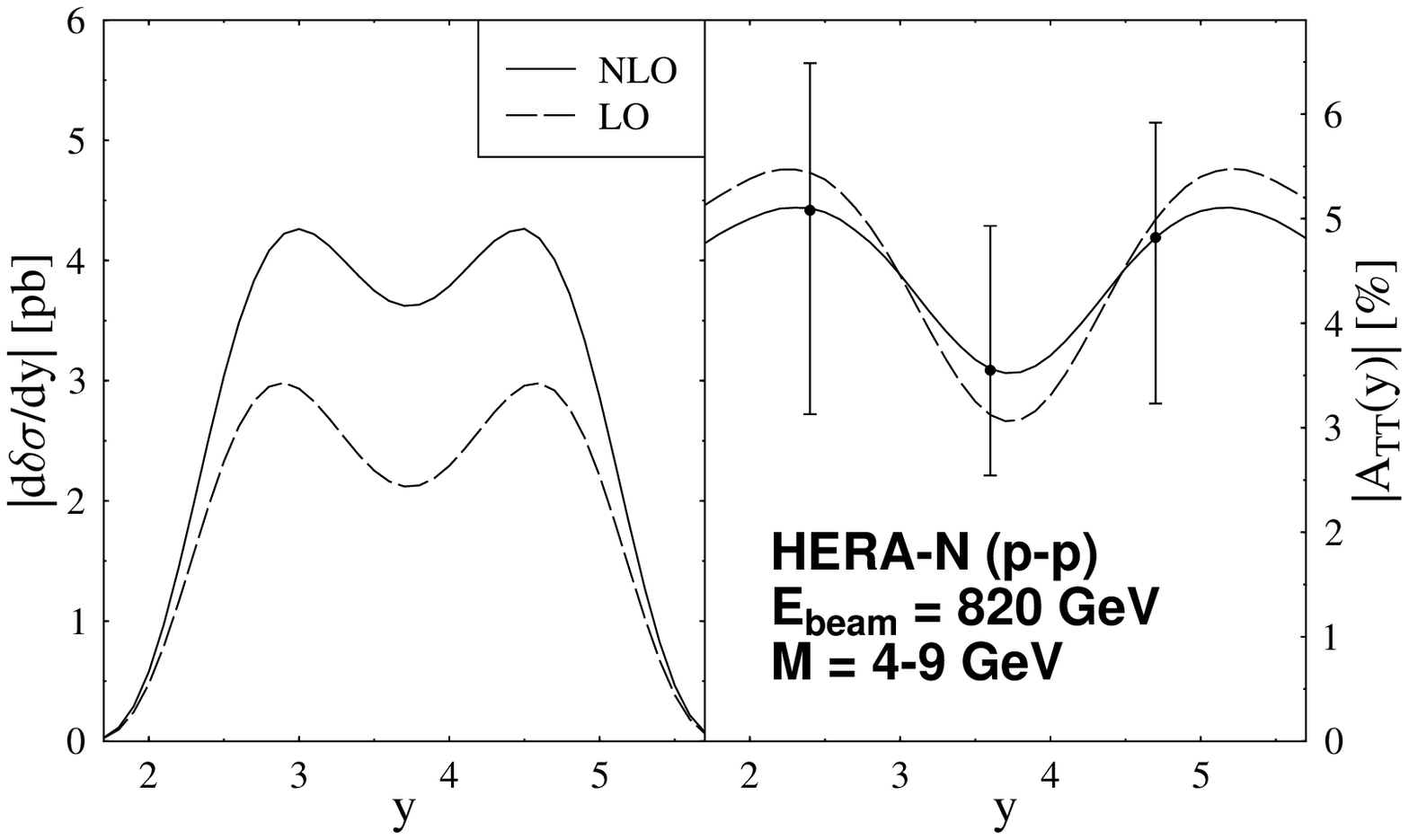,width=15cm}
\large{\bf{Fig.~\ref{fig2}}}

\vspace*{0.8cm}
\epsfig{file=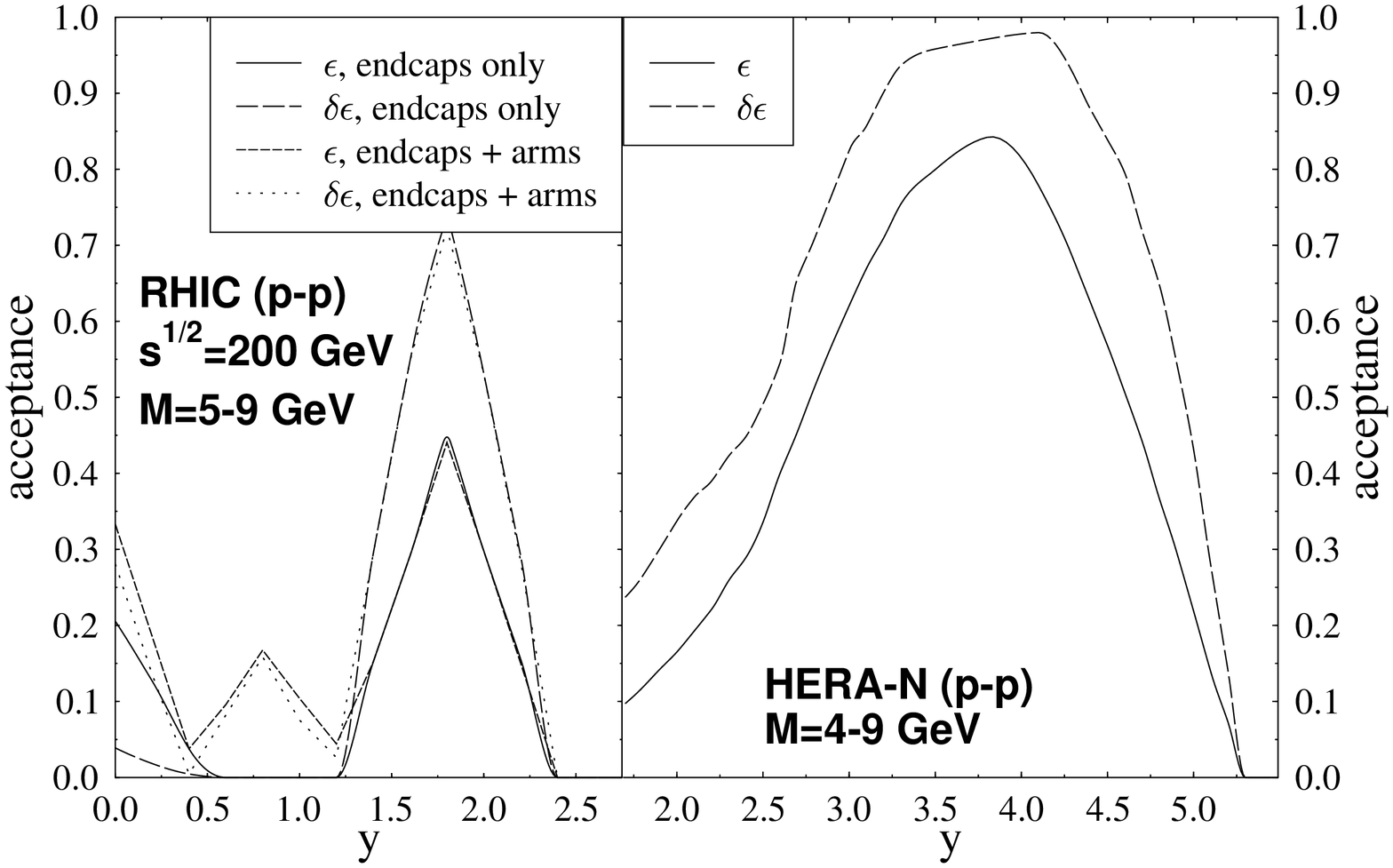,width=14cm}
\large{\bf{Fig.~\ref{fig3}}}

%\vspace*{0.8cm}
\epsfig{file=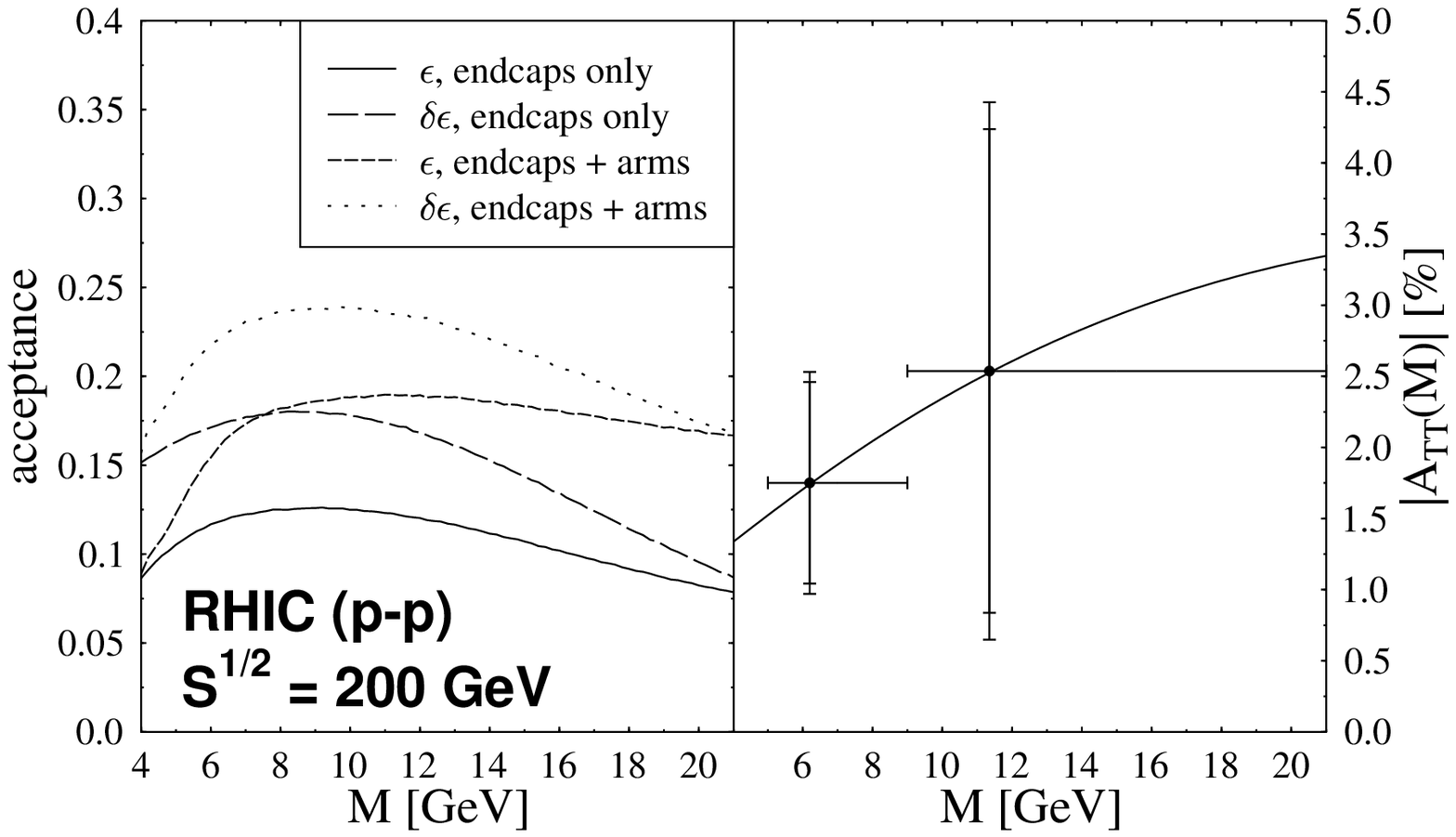,width=15cm}
\large{\bf{Fig.~\ref{fig4}}}

%\vspace*{0.8cm}
%\epsfig{file=fig6.eps,width=15cm}
%\large{\bf{Fig.~\ref{fig5}}}

\vspace*{0.8cm}
\epsfig{file=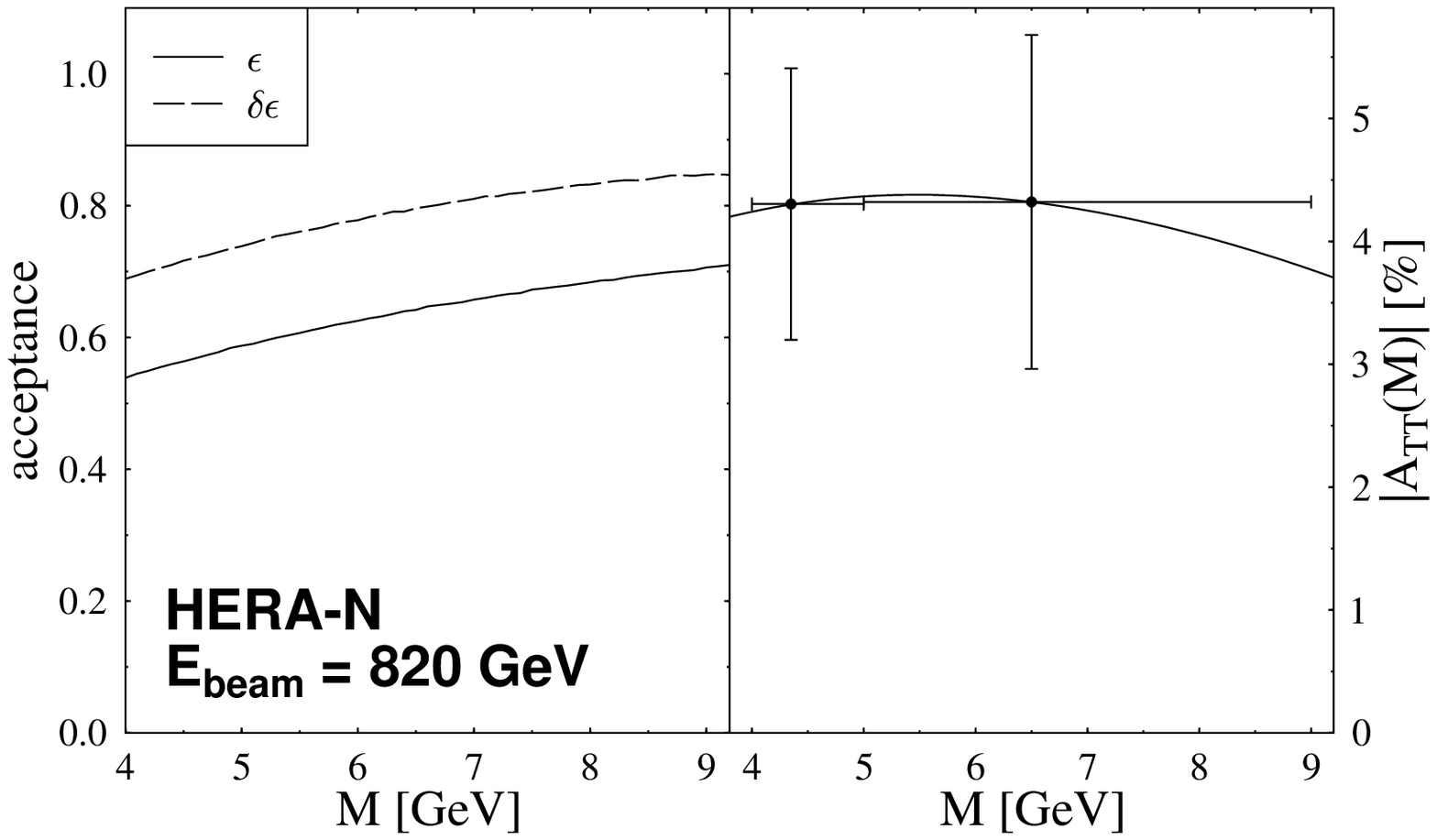,width=15cm}
\large{\bf{Fig.~\ref{fig6}}}

\end{center}

\begin{thebibliography}{99}
\bibitem{rhic} {\sc Rhic} Spin Collab., D.\ Hill et al., Letter of intent 
               {\sc Rhic}-SPIN-LOI-1991, updated 1993;\\
               G. Bunce et al., Particle World {\bf 3}, 1 (1992); \\
               {\sc Phenix}/Spin Collaboration, K.~Imai et al., 
               BNL-PROPOSAL-R5-ADD (1994).
\bibitem{trans} J.P.~Ralston and D.E.\ Soper, Nucl. Phys. {\bf B152}, 
                109 (1979); \\
               R.L.~Jaffe and X.~Ji, Phys. Rev. Lett. {\bf 67}, 552 (1991);
               Nucl. Phys. {\bf B375}, 527 (1992);\\
               X.\ Artru and M.\ Mekhfi, Z. Phys. {\bf C45}, 669 (1990).
\bibitem{soff} J.\ Soffer, Phys. Rev. Lett. {\bf 74}, 1292 (1995).
\bibitem{voge} W.\ Vogelsang, Phys. Rev. {\bf D57}, 1886 (1998).
\bibitem{we}   O. Martin, A. Sch\"afer, M. Stratmann, and W. Vogelsang, 
               Phys. Rev. {\bf D57}, 3084 (1998).
\bibitem{teso} J. Soffer and O. Teryaev, Phys. Lett. {\bf B420}, 375 (1998).
\bibitem{vowe} W.\ Vogelsang and A.\ Weber, Phys. Rev. {\bf D48}, 2073 (1993).
\bibitem{sutton} P. Sutton, A.D. Martin, R.G. Roberts, and
               W.J. Stirling, Phys. Rev. {\bf D45}, 2349 (1992).
\bibitem{kana} Y. Kanazawa, Y. Koike, and N. Nishiyama, 
               Phys. Lett. {\bf B430}, 195 (1998).
\bibitem{saito} N. Saito, private communication.
\bibitem{nowak} V.A. Korotkov and W.-D. Nowak, 
                Nucl. Phys. {\bf A622}, 78c (1997).
\end{thebibliography}
\end{document}